\documentclass[12pt]{elsarticle}

\usepackage{amssymb}
\usepackage{amsthm}

\theoremstyle{definition}
\newtheorem{definition}{Definition}

\usepackage[normalem]{ulem}
\usepackage{graphicx}
\usepackage{tabularx}
\usepackage{url}
\usepackage{tikz}
\usetikzlibrary{positioning}
\usetikzlibrary{patterns}
\usetikzlibrary{shapes}
\usetikzlibrary{arrows}
\usetikzlibrary{calc}
\usetikzlibrary{matrix}

\usepackage[utf8]{inputenc}
\usepackage{pgfplots}

\usepackage{relsize}

\usepackage[numbers]{natbib}
\usepackage{enumerate}
\usepackage{amsmath,amssymb}

\usepackage[plain]{algorithm}
\usepackage[noend]{algpseudocode}
\usepackage{framed}
\usepackage{listings}

\usepackage{afterpage} 

\newcommand{\room}{\hspace{0.8 cm}}

\newcommand{\new}[1]{#1}
\newcommand{\delete}[1]{}

\journal{Parallel Computing}

\begin{document}

\begin{frontmatter}



\title{Bulk-synchronous pseudo-streaming algorithms for many-core accelerators}

\address[cwi]{Centrum Wiskunde \& Informatica, Science Park 123, 1098 XG Amsterdam, The Netherlands.}
\cortext[cor1]{Corresponding author}
\address[qusoft]{QuSoft, Science Park 123, 1098 XG Amsterdam, The Netherlands.}

\author[cwi]{Jan-Willem Buurlage\corref{cor1}}
\ead{j.buurlage@cwi.nl}
\author[cwi,qusoft]{Tom Bannink}
\author[cwi]{Abe Wits}

\begin{abstract}
    The bulk-synchronous parallel (BSP) model provides a framework for writing parallel programs with predictable performance. An important class of parallel hardware in modern high-performance systems are many-core coprocessors. These often have limited locally memory, and can not be targeted by classic BSP programs. In this paper we extend the BSP model by introducing the notion of bulk-synchronous pseudo-streaming (BSPS) algorithms that are able to run on many-core coprocessors. We generalize the BSP cost function to these algorithms, so that it is possible to predict the running time for programs targeting many-core accelerators and to identify possible bottlenecks. To illustrate how to apply the novel framework, two simple examples of BSPS algorithms are explored. To ensure the portability of BSPS software, we propose a small number of additional primitives as an extension to the BSPlib standard. We introduce a software library called Epiphany BSP that implements the introduced concepts specifically for the Parallella development board. Finally, experimental results are given for BSPS algorithms on the Parallella board.
\end{abstract}


\end{frontmatter}



\section{Introduction}
\label{sec:intro}

The \emph{bulk-synchronous parallel} (BSP) model, proposed by Valiant in 1989 \cite{Valiant1990}, is a bridging model for parallel algorithms. The variant of the model we describe here, and use in the remainder of this article, differs slightly from the original model, and will form the basis of the new model that we propose. The BSP computer consists of $p$ processors, assumed to be identical, which each have access to their own local memory. There is also assumed to be a communication network available which can be used by the different processor to communicate with each other. There are bulk synchronizations that ensure that all outstanding communication has been resolved. The time of such a synchronization, the \emph{latency}, is denoted by $l$. The communication cost per data word is denoted by $g$. The parameters $l$ and $g$ are usually expressed in the number of floating-point operations (FLOPs), and are related to \emph{wall time} through the computation rate $r$ of each individual processor which is measured in floating-point operations per second (FLOPS). The four parameters $(p, g, l, r)$ define a BSP computer completely.

A BSP algorithm is structured in a number of supersteps. Each superstep consists of a \emph{computation phase} and a \emph{communication phase}. It is assumed that each processor can simultaneously send and receive data, and that there are no congestion effects in the network, so that the cost of communication is dominated by the maximum number of words sent or received by any processor. At the end of each step a bulk synchronization is performed. Each processor runs the same program, but on different data, which means that BSP algorithms adher to the Single Program Multiple Data (SPMD) paradigm.

Each BSP algorithm has an associated cost, which can be expressed completely using the parameters of a BSP computer. We denote by $w_i^{(s)}$ the amount of work, measured in FLOPs, performed by processor $s$ in the $i$th superstep. We denote by $r_i^{(s)}$ the number of data words received, and with $t_i^{(s)}$ the number of data words transmitted by processor $s$ in superstep $i$. Central to the communication cost of a superstep is the concept of an $h$-relation, which is defined as the maximum number of words transmitted or received by any processor during the superstep, i.e.\ $h_i = \max_{0 \leq s < p} \max \{ t_i^{(s)}, r_i^{(s)} \}$. This leads naturally to the following cost, the BSP cost, of a BSP algorithm consisting of $k$ supersteps:
$$T = \sum_{i = 0}^{k - 1} \left( \max_{0 \leq s < p} w_i^{(s)} + g \, h_i + l  \right).$$

\emph{Streaming algorithms} are a class of algorithms that can be seen as processing methods for sequential data under typically two constraints:
\begin{enumerate}
    \item The computer executing the algorithm has limited (local) memory $L$ available -- typically much less than the total size $S$ of the input, i.e.\ $L \ll S$.
    \item For each part of the input there is only a very limited amount of processing time available (e.g. it is required that the processing should be done in real-time).
\end{enumerate}
Although the main ideas behind streaming algorithms have been studied since the 1980s, the first formal discussion was given in a 1999 article by Alon, Matias and Szegedy \cite{Alon1999}. Many streaming algorithms, in particular because of the second constraint, are massively parallel and often employ randomized methods to provide an approximation (typically called a \emph{sketch}) of the answer. The input of a streaming algorithm takes the form of a \emph{stream} for which we will use the following definition:
\begin{definition}
    A \emph{stream} is an ordered and finite collection of $n$ \emph{tokens}, which we write as
    $$\Sigma = ( \sigma_1, \ldots, \sigma_n ).$$
    Each token is a collection of data that fits in the predetermined local memory size $L$ of the machine processing the stream, i.e.\ the size satisfies $|\sigma_i| \leq L$.
\end{definition}
Many additional constraints can be put on streaming algorithms. For example, they can be enforced to support data streams that can potentially be unbounded in size, or the tokens to be processed are not guaranteed to be presented in any predetermined order, or each token should be discarded or archived after a single pass; see e.g. \cite{Babcock2002}.
In particular, streaming algorithms usually refer to algorithms which only use the input a constant number of times, in many applications even only a single time.

\emph{Many-core coprocessors} are a class of energy-efficient accelerators which focus on massive parallelism. They differ from common multi-core processors found in most desktop computers and servers, by having a large number of simpler processors that typically run at relatively low clock speeds. They can be found in many modern HPC systems\footnote{In the June 2016 TOP-500 supercomputer ranking \cite{TOP500}, the number two supercomputer \textsc{Tianhe-2} uses Intel Xeon Phi coprocessors, while the fastest supercomputer in the world, the \textsc{Sunway TaihuLight}, uses SW26010 many-core processors consisting of 256 cores each.}. Because they are energy-efficient, they are also well suited for use in embedded environments. Algorithms that target these processors are closely related to streaming algorithms, since the constraints that have to be put on the programs strongly resemble the general setting of these algorithms. Furthermore, since the performance of a single processor core is limited, they are optimized for explicit parallel code. Examples of these many-core coprocessors include the Intel Xeon Phi family \cite{Intel}, the Adapteva Epiphany processor \cite{Olofsson2015}, the Movidius Myriad 2 \cite{Movidius} the Kalray MPPA processors \cite{Kalray}, and many others.

We propose a streaming framework within the BSP model which allows BSP algorithms to be generalized so that they can run on these many-core coprocessors with predictable performance.

In the remainder of this article we will discuss this streaming extension to BSP. In Section \ref{sec:bss}, we give a detailed description of this extension. In particular, we introduce the concept of a BSP accelerator and a bulk-synchronous pseudo-streaming algorithm. In Section \ref{sec:examples}, we will give a number of examples of algorithms that fit in this framework. In Section \ref{sec:parallella}, we discuss the Parallella, which is a small parallel computer that will serve as a hardware example to which we can apply the theory that we we introduce in this article. In Section \ref{sec:epiphany}, we discuss the Epiphany processor as a BSP accelerator. Finally, In Section \ref{sec:results} we discuss experimental results for BSPS algorithms obtained with the Parallella board.

\section{Streaming extension to the BSP model}
\label{sec:bss}

Compared to common multi-core processors, many-core coprocessors lack coherent caches, and because of this it is very hard to scale to thousands of processors. Instead, software uses the local memory of each core as a partitoned global address space or as \emph{scratchpad memory}. For explicit parallel programs, the program can only act on the data loaded into the local memory, and therefore an additional layer of software complexity is introduced that is, from the viewpoint of a computer programmer, traditionally handled by the hardware. This layer deals with the data flow between the larger pool of shared memory, and the local memory of each processor core. There is a pressing need for a parallel programming model that can target these systems, while being able to leverage existing algorithms. The approach we take in this article is to use a combination of BSP programs and streaming algorithms to handle this complexity.

\subsection*{BSP accelerator}

We amend the notion of a BSP computer so that it describes more accurately modern many-core hardware, by defining the \emph{BSP accelerator}. In a BSP accelerator, \new{each processing element (in this context called a core)} has limited local memory $L$. \new{In addition, each core} has an asynchronous connection to a \new{shared} external memory pool of size $E \gg L$. This type of connection is commonly available for the many-core coprocessors that motivate this model. We capture the bandwidth to the external memory pool with an additional parameter $e$, the \emph{inverse bandwidth to external memory}, which is defined in FLOPs per data word similar to $g$. A BSP accelerator is completely defined by the parameter pack $(p, r, g, l, e, L, E)$.

In streaming algorithms, both limited local memory as well as limited processing time per data word are assumed. For a BSP accelerator, the processing time need not be limited, but we assume that the local memory is much less then the total input size. We are interested in streaming algorithms not because we want an approximate answer efficiently, but because the amount of local memory of each core of the coprocessor is only sufficient to act on a small part of the input at once.

The input to the algorithms that run on a BSP accelerator is structured into $n$ \emph{streams} $\Sigma^{(i)}$, indexed with $1 \leq i \leq n$. These streams reside in external memory, but can be \emph{opened} by the cores so that they can \emph{stream data down} (read) \emph{and up} (write) from/to a stream. These streams consist of a number of \emph{tokens} $\sigma^{(i)}_j \in \Sigma^{(i)}$, where $1 \leq j \leq |\Sigma^{(i)}|$, and each token fits in the local memory of a core, i.e.\ $|\sigma^{(i)}_j| < L$.

The processing of tokens occurs in a \emph{bulk-synchronous manner}, and is explained in more detail in the upcoming sections. The algorithms we describe here will be written in a SPMD manner, and we assert the completion of the current pass over a token for each processing core before moving on to the next.

Contrary to conventional streaming algorithms, where there is usually a strict order in which the tokens are processed, we are allowed to revisit or skip tokens at any given time. In particular we are free to reuse tokens an arbitrary number of times, and furthermore we assume that we have random access to the tokens within the stream, so that they can be processed in any order. \new{This is similar to \emph{media streaming} where it is possible to \emph{seek} within a stream, allowing one to skip ahead, or revisit arbitrary parts of the audio or video content. After terminology that is common in that context, we will refer algorithms that run on BSP accelerators, which fit into a model that will be described in detail in the next section, as bulk-synchronous \emph{pseudo-}streaming (BSPS) algorithms}.

\subsection*{Hypersteps}

\begin{figure}
\centering
\begin{tikzpicture}

    \draw (0,0) rectangle (9,-2);

    \node at (-0.7, -0.33) {$P(0)$};
    \node at (-0.7, -1.00) {$P(1)$};
    \node at (-0.7, -1.66) {$P(2)$};

    \fill[fill=gray] (0.2, -0.2) rectangle (2.5,-0.5);
    \fill[fill=gray] (0.2, -1.15) rectangle (1.9,-0.85);
    \fill[fill=gray] (0.2, -1.8) rectangle (2.8,-1.5);

    \draw (3,0) rectangle (3,-2);
    \draw (3,0) rectangle (3,-2);

    \fill (3,0) rectangle (3.05,-2);

    \draw[->, thick] (3.2,-0.33) -- (4.2, -1.00);
    \draw[->, thick] (3.2,-0.33) -- (4.2, -1.66);
    \draw[->, thick] (3.2,-1.66) -- (4.2, -0.33);

    \fill[fill=gray, opacity=0.5] (4.4,0) rectangle (4.85,-2);

    \node[rotate=90] at (4.625, -1) {\textsc{sync}};

    \fill[fill=gray] (5.05, -0.2) rectangle (6.05,-0.5);
    \fill[fill=gray] (5.05, -1.15) rectangle (6.50,-0.85);
    \fill[fill=gray] (5.05, -1.8) rectangle (6.95,-1.5);

    \fill (7.15,0) rectangle (7.2,-2);

    \node at (8.075, -1.00) {$\ldots$};

    \draw (4,-2.5) rectangle (10,-6);
    \node[draw,cloud] at (5.5, -3.5) {$\Sigma^{(1)}$};
    \node[draw,cloud] at (7.5, -3.5) {$\Sigma^{(2)}$};
    \node[draw,cloud] at (6.5, -5) {$\Sigma^{(3)}$};
    \node[draw,cloud] at (8.5, -5) {$\Sigma^{(4)}$};
    \node at (4.5, -5.5) {$E$};

    \node at (-0.7, -3.6) {$P(0)$};
    \node at (-0.7, -4.25) {$P(1)$};
    \node at (-0.7, -4.90) {$P(2)$};

    \fill[fill=gray, opacity=0.5] (10.2,0) rectangle (10.65,-6);

    \node[rotate=90] at (10.425, -3) {\textsc{sync}};

    \draw[->, thick] (0,1) -- (10, 1) node [midway, above] {time};

    \draw[<-, dashed, very thick] (0,-3.5) -- (4, -3.5);
    \draw[<-, dashed, very thick] (1,-4.15) -- (4, -4.15);
    \draw[<-, dashed, very thick] (0.5,-4.80) -- (4, -4.80);

    \draw[->, draw=gray, dashed, very thick] (0,-3.7) -- (3, -3.7);
    \draw[->, draw=gray, dashed, very thick] (0,-4.35) -- (3.5, -4.35);
    \draw[->, draw=gray, dashed, very thick] (0,-5.00) -- (2.5, -5.0);
\end{tikzpicture}
\caption{Structure of a hyperstep for $p = 3$ processors and $n = 4$ streams. A hyperstep is a BSP program that is executed on the tokens currently loaded into the cores, together with the concurrent reading of tokens for the next hypersteps. At the end of a hyperstep each core waits until the tokens for the next hyperstep are loaded in for each core, and optionally streams a token containing the results of the hyperstep back up to a stream. Note that the time it takes to fetch the next token may very between cores. Because the fetching of tokens, and the BSP program executed on the current tokens happens concurrently, the total time taken for a hyperstep is dominated by the slowest of the two operations.}
\label{fig:hypersteps}
\end{figure}
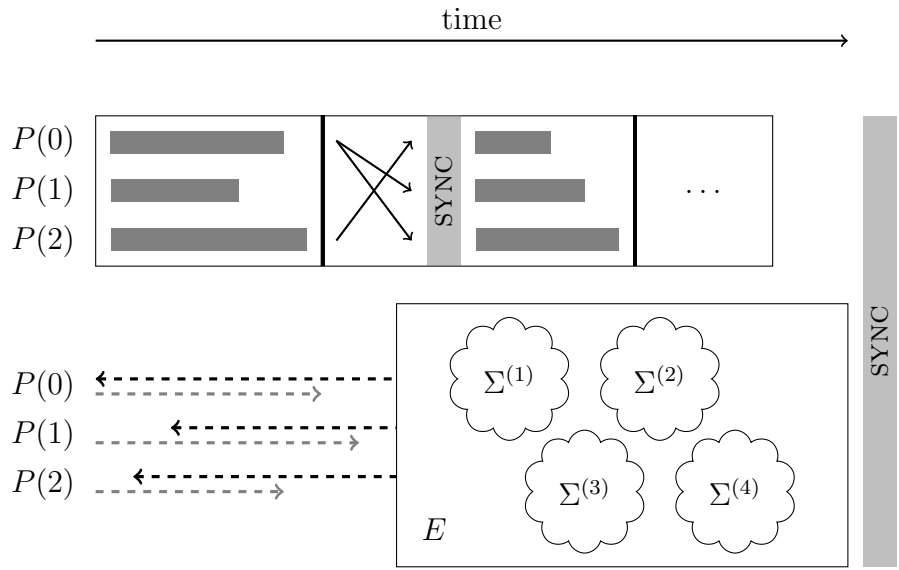

At any given time, the BSP accelerator only has simultaneous access to a limited number of tokens. To cope with this constraint in a systematic manner, we structure the BSPS programs that run on BSP accelerator into a number of \emph{hypersteps}. A hyperstep consists of two operations; 1) an (ordinary) BSP program that is performed on the tokens that are currently loaded into the cores, and 2) the \emph{fetching} of tokens that will be used in the next superstep. Although the computations on a core are limited to the tokens that it has available in its local memory, it can communicate with the other cores of the BSP accelerator. After such a step, there is a global bulk-synchronization before every \new{core} moves on to the next hyperstep. This ensures that each core has all the tokens required for the next hyperstep available in its local memory. See also Figure \ref{fig:hypersteps}. The data streams are prepared by an external processing unit which we will call \emph{the host}, and for our purposes it is considered a black box.

Since we assume that there is an asynchronous communication mechanism with the external memory pool, obtaining the tokens for the next hyperstep \emph{(prefetching)} can be done concurrently with the BSP program of the current hyperstep. In practice, the next tokens are written to a local buffer so that after a hyperstep the next hyperstep can be initiated as soon as possible. This minimizes the down-time within hypersteps, and is reminiscent of cache prefetching techniques. Note that prefetching data halves the effective local memory size, since storage needs to be reserved for the buffer that holds the next token.

\subsection*{BSPS cost}

Next we want to generalize the BSP cost, and define the BSPS program consisting of $H$ hypersteps.  Each hyperstep $0 \leq h < H$ has an associated BSP program, with a BSP cost $T_h$. In every hyperstep (except for the last, which is a technicality we will ignore) the next tokens are prefetched from the external memory $E$ with inverse bandwidth $e$. In our discussion \new{we will set the size of a data word to be equal to the size of a \emph{floating point number}. We will therefore allow ourselves to write `float' when talking about message sizes and buffers.} For simplicity we assume that the \new{tokens of the $i$th stream have constant size $C_i$}, and furthermore we assume that the first tokens are available for each core of the accelerator at the start of the program. The set of indices of the \emph{active streams} of core $s$, i.e.\ the streams from which tokens are being read by the core, we will denote with $\mathcal{O}_s$. The BSPS cost of a single hyperstep then corresponds to the maximum between the time spent processing the current tokens, denoted by $T_h$, and the time taken to fetch the next tokens. This leads to a natural definition for the cost function of a BSPS program:
\begin{equation}
    \tilde{T} = \sum_{h = 0}^{H - 1} \max\left(T_h, e \max_{0 \leq s < p} \sum_{i \in \mathcal{O}_s} C_i\right).
    \label{eqn:bspscost}
\end{equation}
If fetching the next token takes more time than processing the current token, then the running time of the hyperstep is bound by the memory bandwidth. If this happens, we say that the hyperstep is \emph{bandwidth heavy}. Otherwise we say that the hyperstep is \emph{computation heavy}.

\new{The advantages of this pseudo-streaming paradigm for programming many-core processors include:
\begin{itemize}
    \item Streams and tokens \emph{guide} algorithm designers and implementers to use predictable and efficient access patterns when dealing with external data.
    \item The resulting algorithms are amenable to \emph{precise run-time analysis}.
    \item The model provides a \emph{simple way to describe and implement portable parallel algorithms for many-core accelerators}, and the library implementation we describe minimizes the necessary boilerplate.
\end{itemize}
We believe these advantages will ultimately lead to performant parallel algorithms and programs that will run across a variety of modern platforms.}

\subsection*{\new{Comparison with previous work}}

The main contribution of this paper is the new pseudo-streaming programming model that provides a convenient and portable way to develop algorithms for many-core accelerators. As we will show, a powerful feature of this paradigm is that existing BSP algorithms can be reused within hypersteps of BSPS algorithms, so that the programmer often only needs to worry about the communication with the shared external memory, for which streaming algorithms and techniques can be employed. This powerful interplay between streaming algorithms and BSP algorithms lead to elegant implementations of algorithms for many-core coprocessors.

The main distinction between the classic BSP performance model and the model we consider here, is the asynchronous fetching from external memory. Extensions to the BSP model that specify parameters for the memory size have already been studied before, see e.g. \cite{McColl1999, Tiskin1998}. In particular, there exists an external memory extension to the BSP model, EM-BSP \cite{Dehne2003}, and many algorithms have been considered in this context \cite{Dehne2002}. In the EM-BSP model, each core has a synchronous connection to secondary local memories. Instead, we consider a single shared external memory pool, and algorithms optimized for limited local memory. The performance model we describe here is similar to the multi-memory BSP (MBSP) model \cite{Gerbessiotis2015}, which supports multiple units of external memory. Compared to MBSP, the BSPS model has a simplified view of the accessible memory; since it only distuinguishes between external and local memory, and has the important advantage of providing an explicit programming model that is well adapted to modern many-core accelerators and that can leverage ideas from BSP algorithms as well as from the large body of streaming algorithms that exist.

Another recent development is Multi-BSP \cite{Valiant2011}, introduced as a model for writing portable parallel algorithms for general multi-core systems. The Multi-BSP model uses a tree structure to represent the memory hierarchy of a computer. In our view, this model is overly complicated for the architectures that we target with this work, and the explicit asynchronous prefetching we discuss can not be incorporated within the model.

Finally we note that although here we exclusively use BSP as our on-core model for the parallel programs that run during a hyperstep, there is a lot of flexibility in this choice, other on-core models can easily be incorporated into the BSPS cost function. We mention for example D-BSP \cite{Bilardi2001}, which allows for varying parameters between different clusters of cores, and which may become relevant as the on-core model as the number of cores on a single chip increase.

\section{Examples}
\label{sec:examples}

In this section we will discuss two simple examples of algorithms that fit into the framework we described. First we will discuss BSPS algorithms for computing the inner-product of two vectors, and for performing the matrix multiplication of two dense matrices. 

We use the following functions, with their usual semantics, in the descriptions of the algorithms:
\begin{table}[H]
\centering
\begin{tabular}{| l | l |}
\hline
    $\sigma \gets \textsc{read}(\Sigma)$ & read token $\sigma$ from its stream \\
    \textsc{write($\sigma$, $\Sigma$)} & write token $\sigma$ to stream $\Sigma$ \\
    \textsc{broadcast}($a$) & send the value $a$ to all other cores \\
\textsc{sync} & perform a bulk-synchronization of all the cores \\
    \textsc{move}($\Sigma, k$) & change the next token read off of $\Sigma$ by $k$ tokens. \\
\hline
\end{tabular}
\end{table}

\subsection{Inner-product}

As a simple example to illustrate the main BSPS concepts, we will first consider the inner product of two vectors $\vec{v}, \vec{u} \in \mathbb{R}^N$ of size $N$, and construct a BSPS algorithm to compute
$$\alpha = \vec{v} \cdot \vec{u} = \sum_{i = 1}^N v_i u_i.$$
Here, we assume that the total number of components $v_i$ that can be stored at a single core is much smaller than the total size of the vector.

\begin{figure}
\centering
\begin{tikzpicture}[scale=0.8]
    \node at (-0.75, 0.5) {$\vec{v}$};
    \filldraw[fill=white] (0, 0) rectangle (5, 1);
    \filldraw[fill=blue] (0, 0) rectangle (1, 1);
    \filldraw[fill=red] (1, 0) rectangle (2, 1);
    \filldraw[fill=green] (2, 0) rectangle (3, 1);
    \filldraw[fill=blue] (3, 0) rectangle (4, 1);
    \filldraw[fill=red] (4, 0) rectangle (5, 1);
    \filldraw[fill=green] (5, 0) rectangle (6, 1);
    \filldraw[fill=blue] (6, 0) rectangle (7, 1);
    \filldraw[fill=red] (7, 0) rectangle (8, 1);
    \filldraw[fill=green] (8, 0) rectangle (9, 1);
    \filldraw[fill=blue] (9, 0) rectangle (10, 1);
    \filldraw[fill=red] (10, 0) rectangle (11, 1);
    \filldraw[fill=green] (11, 0) rectangle (12, 1);

\begin{scope}[xshift=10.5em]
    \node at (-0.75, -1.5) {$\Sigma^{\vec{v}}_0$};
    \filldraw[fill=blue] (0, -2) rectangle (1, -1);
    \filldraw[fill=blue] (1, -2) rectangle (2, -1);
    \filldraw[fill=blue] (2, -2) rectangle (3, -1);
    \filldraw[fill=blue] (3, -2) rectangle (4, -1);
    \draw[thick] (2, -2.5) -- (2, -0.5);
    \node at (1, -2.6) {$\sigma^{\vec{v}}_1$};
    \node at (3, -2.6) {$\sigma^{\vec{v}}_2$};
\end{scope}

\end{tikzpicture}
\caption{Here we depict the construction of the streams used in the inner-product problem. The stream shown here is $\Sigma^{\vec{v}}_0$, corresponding to the components of $\vec{v}$ assigned to the first processor. We consider $p = 3$ processors in total. Each token consists of $C = 2$ vector components, and the total stream size is $|\Sigma_0^{\vec{v}}| = 4$.}
\label{fig:innerproductstream}
\end{figure}
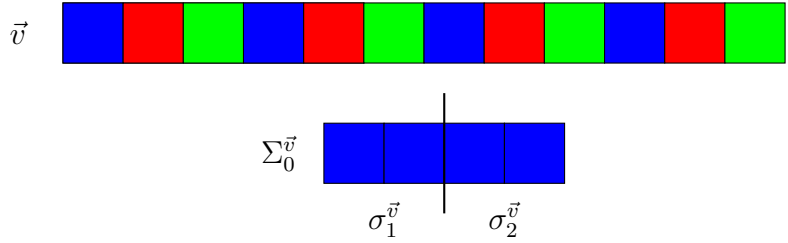

We begin by implicitly distributing the vectors over the processing cores of our BSP accelerator. In this discussion we will use a cyclic distribution of the vector so that $v_i$ and $u_i$ are assigned to the processor with index $s = i \bmod p$. Next we partition the resulting data for the $s$th core, which we will take as the streams $\Sigma^{\vec{v}}_s$ and $\Sigma^{\vec{u}}_s$, into a number of tokens, each of which will fit in a designated chunk of local memory with a certain \emph{token size} $C$, see also Figure \ref{fig:innerproductstream}.

\begin{algorithm}
\begin{framed}
    \begin{algorithmic}
        \State \textbf{Input: } $\Sigma^{\vec{v}}_s = \{ \sigma_1^{\vec{v}}, \sigma_2^{\vec{v}}, \ldots, \sigma_n^{\vec{v}} \}$, $\Sigma^{\vec{u}}_s = \{ \sigma_1^{\vec{u}}, \sigma_2^{\vec{u}}, \ldots, \sigma_n^{\vec{u}} \}$
        \State \textbf{Output: } $\alpha = \vec{v} \cdot \vec{w}$
    \\\State $\alpha_s \gets 0$
    \For{$1 \leq i \leq n$}
        \State $\sigma_i^{\vec{v}} \gets \textsc{read}(\Sigma^{\vec{v}}_s)$
        \State $\sigma_i^{\vec{u}} \gets \textsc{read}(\Sigma^{\vec{u}}_s)$
    \State $\alpha_s \gets \alpha_s + \sigma_i^{\vec{v}} \cdot \sigma_i^{\vec{u}}$
    \EndFor
    \\
        \State \textsc{broadcast}($\alpha_s$)
        \State \textsc{sync}
        \State $\alpha \gets \sum_{t = 0}^{p - 1} \alpha_t$
    \end{algorithmic}
    \end{framed}
    \caption{Summary of the BSPS algorithm for computing the inner product. After the completion of the algorithm every core of the accelerator will have computed the value $\alpha = \vec{v} \cdot \vec{u}$. This value can then be communicated back to the host.}
    \label{alg:bspsinnerproduct}
\end{algorithm}

Every core maintains a partial sum $\alpha_s$ throughout the algorithm. We consider each pair of tokens (both consisting of $C$ vector components) and compute locally the inner product of this subvector and add it to $\alpha_s$. After every token has been considered, the combined partial sums of all the cores will be equal to the desired value for the inner product $\alpha$. Note that we can identify a token with a subvector, and we construct the streams for the two vectors in a completely identical manner. We summarize the algorithm in Algorithm \ref{alg:bspsinnerproduct}.

Let us consider the BSPS cost of this algorithm. The total number of hypersteps is equal to $n = \frac{N}{pC}$. The last hyperstep is followed by an ordinary superstep in which the sum of partial sums is computed, where each processor sends and receives $(p - 1)$ data words. In each of these hypersteps we compute an inner product between two vectors of size $C$, taking $2C$ time, and this requires no communication. The total BSPS cost of this algorithm is:
$$T_{\text{inprod}} = n \cdot \max \{ 2C, 2C \, e \} + p + (p - 1)g + l.$$
We see that if $e > 1$ then the hypersteps are bandwidth heavy, otherwise they are computation heavy.

\subsection{Dense matrix-matrix multiplication}

The next algorithm we consider in this context is the product of two dense matrices, which are too large to fit completely in the local memory of the accelerator. The resulting algorithm will be an adaptation of Cannon's algorithm \cite{Cannon1969} which computes this product on a square grid of accelerator cores.

\subsubsection*{Cannon's algorithm}

We first describe Cannon's algorithm. We want to compute $AB = C$ for two matrices $A$ and $B$, and assume we have $N \times N$ processors. We index each processor core with a pair $(s, t)$. The matrices $A$, $B$ and $C$ are split into $N \times N$ blocks of equal size (padding with zeros if necessary):
$$A = \left( \begin{array}{c | c | c | c} A_{11} & A_{12} & \ldots & A_{1N} \\ \hline A_{21} & A_{22} & \ldots & A_{2N} \\ \hline \vdots & \vdots & \ddots & \vdots \\ \hline A_{N1} & A_{N2} & \ldots & A_{NN}\end{array} \right),$$
so that we can write for the resulting blocks of $C$:
$$C_{ij} = \sum_{k = 1}^N A_{ik} B_{kj} \room 1 \leq i, j \leq N.$$
We see that the resulting block $C_{ij}$ is the result of adding $N$ terms, in each of which a block of $A$ and a block of $B$ are multiplied. Since there are exactly $N \times N$ blocks $C_{ij}$, and $N \times N$ processors, it would be very natural to let each processor compute exactly one of these sums in $N$ steps. However there is one immediate problem: many blocks of $A$ and $B$ are needed simultaneously in the same step $k$, and we do not want to copy our blocks to every single processor since we assume that there is finite storage, and therefore limited room for duplication. Luckily, the sum above can be rearranged so that in step $k$ the processor requiring a specific block of $A$ or $B$ is unique, so that we never require any data redundancies. After computing a term, the matrix blocks that were used can be moved around to the processor that needs the block next. We let the processor with index $(s, t)$ compute the product:
$$A_{s, 1 + (t + s + k - 3) \bmod N} B_{1 + (s + t + k - 3) \bmod N, t}$$
in the $k$th step\footnote{Note that the indices would be much more straightforward had we used $0$-based indices for our matrices, but we will stick with $1$-based indices in this discussion for consistency.}. Next, we consider which processor needs the blocks of the current step after a processor is done with it. In the $(k+1)$th step, the processor $(s, t)$ needs the $A$ block that was previously owned by processor $(1 + (s + t + k - 3 \bmod N), t)$ in the previous step, while the $B$ block was previously owned by $(s, 1 + (s + t + k - 3 \bmod N))$. In summary, we have the following scheme:
\begin{enumerate}
    \item Perform an initial distribution of the matrix blocks over the $N \times N$ processors, sending
        $A_{i, j} \mapsto (i, 1 + ((i + j - 2) \bmod N))$ and $B_{i, j} \mapsto (1 + ((i + j - 2) \bmod N), j)$.
\item Let each processor compute the product of the two local matrix blocks of $A$ and $B$, adding the result to $C_{st}$.
\item Next each processor sends the matrix block of $A$ to the right, i.e. to processor $(s, 1 + (t \bmod N))$, and each matrix block of $B$ down to processor $(1 + (s \bmod N), t)$. We repeat steps 2 and 3 a total number of $N$ times.
\end{enumerate}
The resulting matrix product $C$ will then be available distributed over the processors. We wil refer to this algorithm in the program text as $\textsc{cannon}$.

\subsubsection*{Multi-level Cannon's algorithm}

We will now generalize this algorithm to a BSPS variant. The method we discuss here is similar to the one described in e.g. \cite{Ross2015}. The distribution scheme that we derived in the previous section will not suffice in general for BSP accelerators, since for reasonably large matrices, dividing them into $N \times N$ blocks will not make them small enough so that they can be stored in the local memory of the cores. Thus, we need to reduce these sub-problems in size even further. We do this by subdividing the matrix in two levels. The first level will no longer consist of $N \times N$ blocks, but of $M \times M$ blocks, where $M$ is taken suitably large. Each of these blocks will be divided further in $N \times N$ blocks, which will be distributed over the cores in the method described above.

For example $A$ will now look like this:
$$A = \left( \begin{array}{c | c | c | c} A_{11} & A_{12} & \ldots & A_{1M} \\ \hline A_{21} & A_{22} & \ldots & A_{2M} \\ \hline \vdots & \vdots & \ddots & \vdots \\ \hline A_{M1} & A_{M2} & \ldots & A_{MM}\end{array} \right),$$
where each outer block $A_{ij}$ is divided further as:
$$A_{ij} = \left( \begin{array}{c | c | c | c} (A_{ij})_{11} & (A_{ij})_{12} & \ldots & (A_{ij})_{1N} \\ \hline (A_{ij})_{21} & (A_{ij})_{22} & \ldots & (A_{ij})_{2N} \\ \hline \vdots & \vdots & \ddots & \vdots \\ \hline (A_{ij})_{N1} & (A_{ij})_{N2} & \ldots & (A_{ij})_{NN}\end{array} \right).$$
In total we then have $MN \times MN$ blocks. We can choose our value of $M$ such that the resulting smaller blocks $(A_{ij})_{kl}$ are small enough to fit in the local memory of the cores.
Let us now turn our attention to constructing the streams. We will consider the $M^2$ blocks of $A$ in row-major order, and the $M^2$ blocks of $B$ in column-major order. The blocks will form the tokens, and will all be considered $M$ times. In every hyperstep we will compute the product of two blocks using Cannon's algorithm introduced above. To construct the stream, we will denote with e.g.\ $(A_{ij})_{st}$ the first inner block that the processor $(s, t)$ receives when computing a product involving the outer block $A_{ij}$. We define $(B_{ij})_{st}$ in a similar manner. We are now ready to define the streams:
\begin{align*}
    \Sigma^A_{st} = &\underset{\circlearrowright~M~\text{times}}{\underbrace{(A_{11})_{st} (A_{12})_{st} \ldots (A_{1M})_{st}}}~\underset{\circlearrowright~M~\text{times}}{\underbrace{(A_{21})_{st} (A_{22})_{st} \ldots (A_{2M})_{st}}} \\
&\ldots \underset{\circlearrowright~M~\text{times}}{\underbrace{(A_{M1})_{st} (A_{M2})_{st} \ldots (A_{MM})_{st}}},
\end{align*}
and
\begin{align*}
    \Sigma^B_{st} = &(B_{11})_{st} (B_{21})_{st} \ldots (B_{M1})_{st} (B_{12})_{st} (B_{22})_{st} \\
&\underset{\circlearrowright~M~\text{times}}{\underbrace{\ldots (B_{M2})_{st} (B_{13})_{st} \ldots (B_{1M})_{st} (B_{2M})_{st} \ldots (B_{MM})_{st}}}.
\end{align*}
Here, we indicate with $\circlearrowright$ the order in which we consider the tokens, so that $\circlearrowright M$ means that we will repeat looping over that particular section of blocks $M$ times before moving on to the next section of blocks. Note that each block is only stored in the stream once. We will loop over groups of $M$ blocks of $A$ a number of $M$ times before moving to the next, while we simply loop over the $M^2$ blocks of $B$ a total number of $M$ times.
After constructing these streams, from the perspective of an accelerator we have to multiply the two tokens, corresponding to the outer matrix blocks, given to us in each of the $M^3$ hypersteps. This is done by computing the product of the two blocks with the general Cannon's algorithm, which can now be applied since we have chosen the outer blocks to be of small enough size. The result of this product is added to the block $C_{ij}$ that is currently being computed. After every $M$ hypersteps we have completely computed one of the $M^2$ blocks of $C$, and we store the result in the external memory $E$.

\begin{algorithm}
\begin{framed}
    \begin{algorithmic}
        \State \textbf{Input: } $\Sigma^{A}_{st}, \Sigma^{B}_{st}$, $M$
        \State \textbf{Output: } $\Sigma^C_{st}$.
        \\
    \For{$1 \leq i \leq M$}
    \For{$1 \leq j \leq M$}
        \State $\sigma^{C}_{ij} \gets \vec{0}$
        \For{$1 \leq k \leq M$}
            \State $\sigma^A \gets \textsc{read}(\Sigma^{A}_{st})$
            \State $\sigma^B \gets \textsc{read}(\Sigma^{B}_{st})$
            \State $\textsc{cannon}(\sigma^B, \sigma^B, \sigma^{C}_{ij})$
        \EndFor
        \State $\textsc{write}(\sigma^{C}_{ij}, \Sigma^C_{st})$
        \State $\textsc{move}(\Sigma^A_{st}, -M)$
    \EndFor
        \State $\textsc{move}(\Sigma^B_{st}, -M^2)$
    \EndFor
    \\
    \end{algorithmic}
    \end{framed}
    \caption{Summary of the BSPS version of Cannon's algorithm that runs on core $(s, t)$. Here, $\vec{0}$ denotes an array of zeros.}
    \label{alg:bspscannon}
\end{algorithm}

Let us consider the BSPS cost of this algorithm. First we will derive the BSP cost of Cannon's algorithm. There are $N$ supersteps in which we compute the product of two inner blocks of size $k \times k \equiv \frac{n}{NM} \times \frac{n}{NM}$, which takes $2 k^3$ flops. Next we send and receive such an inner block consisting of $k^2$ words. Note that we do not send or receive such a block in the final superstep, but for simplicity we will ignore this. The BSP cost equals:
$$T_{\text{cannon}} = N ( 2 k^3 + k^2 g + l).$$
The number of values in a token, the token size $C$, is given by the number of values in an inner block which is equal to $k^2$. For simplicity, we will ignore the costs of storing the resulting blocks. There are $M^3$ hypersteps, so that we can write for the BSPS cost of this algorithm:
\begin{equation}
    \tilde{T}_{\text{cannon}} = M^3(\max(N ( 2 k^3 + 2 k^2 g + l), 2 k^2 e)),
    \label{eq:cannontime} 
\end{equation}
Alternatively, after substituting back $k$, we can write for the total cost:
\begin{equation*}
     \tilde{T}_{\text{cannon}} = \max \left( 2 \frac{n^3}{N^2} + \frac{2 M n^2}{N} g + N M^3 l,~2\frac{M n^2}{N^2} e \right).
\end{equation*}

\section{Streaming extension to BSPlib}
\label{sec:parallella}

\lstset{
  belowcaptionskip=1\baselineskip,
  language=C,
  tabsize=4,
  basicstyle=\footnotesize\ttfamily,
  keywordstyle=\bfseries\color{blue!80!black},
  identifierstyle=\color{black},
}

In this section we propose a streaming extension to the BSPlib standard \cite{Hill1998}. This initial proposal targets simple streaming applications. The BSPlib standard has been extended previously with high-performance primitives \cite{Yzelman2014}, which we have adopted. In addition, we introduce a number of new primitives that can be used to write BSPS programs. A BSPS program consists of a \emph{host program} that runs on the host, and a \emph{kernel} that runs on the cores of the accelerator. For a more detailed specification of these new primitives we refer to the Epiphany BSP documentation \cite{Buurlage2015}, which is a software package that provides implementations for the primitives introduced here.

The host of a BSP accelerator needs to be able to create streams of data. To create a stream we have to specify respectively the total size, the size of the tokens, and optionally it is possible to set the initial data of the stream.

\begin{lstlisting}
void* bsp_stream_create(int stream_size, int token_size,
                        const void* initial_data);
\end{lstlisting}

This is the only new primitive that is called from the host. The return value is a pointer to a buffer for the data in the stream. Streams are given an identifier \texttt{stream\_id} in order of creation, starting from index $0$ and increasing by one each time a stream is created. Inside a kernel program, streams can be opened and closed using the following primitives:

\begin{lstlisting}
int bsp_stream_open(bsp_stream* stream, int stream_id);
int bsp_stream_close(bsp_stream* stream);
\end{lstlisting}

Here, \texttt{bsp\_stream} is a C struct that holds the required information for a stream.

Streams are shared between cores. Streams can only be opened if they are not yet opened by another core. After opening a stream, tokens can be obtained from it. After closing the stream any core can open it again. The return value is equal to the maximum size of a token in bytes.

Tokens can be moved down from open streams. Furthermore, data can be streamed back up to the streams, i.e.\ the streams are mutable. For this the following primitives are used:

\begin{lstlisting}
int bsp_stream_move_down(bsp_stream* stream, void** buffer,
                         int preload);
int bsp_stream_move_up(bsp_stream* stream, const void* data,
                       int data_size, int wait_for_completion);
\end{lstlisting}

These functions return the size in bytes of the buffer that will hold the next token. The location of this buffer is written to \texttt{*buffer}. The argument \texttt{preload} should be set to either $0$ or $1$, this respectively disables, or enables \emph{prefetching} (see Section \ref{sec:bss}) the next token. The parameters of the second function are self-explanatory.

It is possible to (re)use a token at different stages of your algorithm. A cursor is maintained for each stream which corresponds to the next token that should be obtained or written to. This cursor can be modified using the following primitive:

\begin{lstlisting}
void bsp_stream_seek(bsp_stream* stream, int delta_tokens);
\end{lstlisting}

Here \texttt{delta\_tokens} denotes the number of tokens the cursor should move, relative to the current position. Note that this mechanism gives us random access inside the streams.

\section{The Epiphany processor as a BSP accelerator}
\label{sec:epiphany}

The Parallella\footnote{Parallella: a supercomputer for everyone. \url{https://www.kickstarter.com/projects/adapteva/parallella-a-supercomputer-for-everyone}} is a ``credit card-sized computer'' intended to make parallel programming accessible and open to a large community. It is similar to other small-form computing platforms such as the popular Raspberry Pi\footnote{Rasberry Pi: a low cost, credit-card sized computer: \url{https://www.raspberrypi.org/}} and Arduino\footnote{Arduino: an open-source electronics platform \url{https://www.arduino.cc/}}.

The Parallella board has basic network capabilities and support for a number of peripherals. There are two different processors available. The \emph{host} processor, which runs the (Linux) operating system, is a dual-core ARM processor. The \emph{coprocessor} on the Parallella board is based on the Epiphany architecture which has 16 RISC (Reduced Instruction Set Computer) cores.
The Epiphany processor architecture \cite{Adapteva2013} defines a square grid of cores of size $N \times N$. On the processor there is also a network-on-chip (NOC) present. There is support for single-precision floating point operations. The chip supports core-to-core communication on the processor with very low latency (in the order of nanoseconds) and zero start-up costs. Besides the 16 core Epiphany-III processor, there has also been a limited production of Parallella boards with the Epiphany IV processor which has 64 cores. The Epiphany-V coprocessor has recently been announced. Although it is not yet available, it will have 1024 cores and has support for double-precision operations \cite{Olofsson}.
\begin{figure}
\centering
\begin{tikzpicture}
    \tikzstyle{snode}=[thick, shape=rectangle, inner sep=0.3 cm, text opacity=1]
    \tikzstyle{pnode}=[draw=black]
    \tikzstyle{mnode}=[draw=gray]
    \node[draw, snode, pnode] (Host) {Host};
    \node[draw, snode, mnode, right=of Host] (RAM) {main};
    \node[draw, snode, mnode, left=of Host] (DRAM) {shared};
    \node[draw, snode, pnode, left=of DRAM] (Epiphany) {Epiphany};
    \node[draw, snode, mnode, left=of Epiphany] (SRAM) {local};
    \draw[<->, thick] (Host) -- (RAM) node [midway, above, sloped] {};
    \draw[<->, thick] (Host) -- (DRAM) node [midway, above, sloped] {slow};
    \draw[<->, thick] (Epiphany) -- (DRAM) node [midway, above, sloped] {slow};
    \draw[<->, thick] (Epiphany) -- (SRAM) node [midway, above, sloped] {fast};
\end{tikzpicture}
\caption{Overview of the Parallella memory. There are three kinds of memory: (A) 32 kB local memory per core, (B) 32 MB of shared DRAM, (C) 1 GB of main DRAM. We also give an indication of the relative speed of the different memory lanes that are available. There is also a slow connection between the host and the local memory of each core which we do not use and will therefore ignore.}
\label{fig:parallellamemory}
\end{figure}
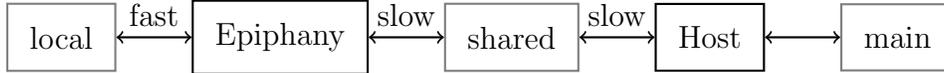
We distinguish bewteen three layers of memory on the Parallella board. There is 1GB of RAM available, which is split into two parts; The largest part is exclusive to the host processor, and we will simply refer to it as RAM. A relatively small section is shared between the host and the Epiphany, and is called the DRAM (or \emph{dynamic} memory). Finally, there is 32 kB of local memory present at each core, which we will refer to as the SRAM (or \emph{static} memory). See Figure \ref{fig:parallellamemory}. An important feature is the availability of two direct memory access (DMA) engines at each Epiphany core. These allow for asynchronous reading and writing between Epiphany cores, and between the (local memory of) Epiphany cores and the dynamic memory, and will play an important role in our implementation of pseudo-streaming algorithms.

\subsection{Epiphany BSP}

We developed Epiphany BSP \cite{Buurlage2015} (EBSP) as an implementation of the BSPlib standard \cite{Hill1998} on top of the Epiphany SDK provided for the Parallella. It is released under the lesser GNU public license (LGPL). Other libraries and technologies that are supported on the Parallella include MPI \cite{Ross2015}, OpenMP \cite{Agathos2015}, Erlang \cite{Fleming2015}, and OpenCL \cite{BrownDeerTechnology2014}.

A typical Epiphany BSP application consists of two separate programs. The \emph{host program} configures the application and prepares the data to be processed by the Epiphany coprocessor. The \emph{kernel} is a program that runs on each of the Epiphany cores in a SPMD manner. All the communication between Epiphany cores, and between the host and the coprocessor can be done using the conventional BSP methods and syntax (e.g. buffered and unbuffered writes or through message passing mechanisms). A major goal of the development of EBSP is to allow current BSP programs to be run on dual-processor hardware such as the Parallella with minimal modifications.

The Epiphany BSP library also provides many utilities to ease the development of BSP applications for the Parallella board, such as timers, dynamic memory management and debugging capabilities. Finally, EBSP also provides an extension to BSP to support streaming algorithms. We will introduce and formalize this extension in the next section.

We will consider the Epiphany-III 16-core Microprocessor (E16G301) chip that is found on the original Parallella board as a concrete example of a BSP accelerator.  As we mentioned when we introduced the Parallella in Section \ref{sec:parallella}, the Epiphany chip is connected to a portion of memory called the DRAM which we will take as our external memory $E$, and each core comes equipped with a DMA engine which gives us an asynchronous connection to this memory pool.
There are many possible communication paths between the host, the Epiphany and the various kinds of memory. We are interested in estimating as accurately as possible the inter-core communication speed $g$, the latency $l$, and the read/write speed $e$ from an Epiphany core to the external memory using the DMA engine.
\begin{table}
    \caption{The communication speeds \new{to shared memory} that were obtained from measurements done during the development of Epiphany BSP. In the \emph{network state} column we indicate if a single core is reading/writing (free) or if all cores are reading/writing simultaneously (contested). All the speeds are given \emph{per core}.}
\label{tab:benchmarkparallella}
\begin{tabularx}{\textwidth}{|X|X|X|X|}
    \hline
    \textbf{Actor} & \textbf{Network state} & \textbf{Read} & \textbf{Write} \\
    \hline
    Core & contested & 8.3 MB/s & 14.1 MB/s\\
         & free & 8.9 MB/s & 270 MB/s\\
    \hline
    DMA & contested & 11.0 MB/s & 12.1 MB/s\\
        & free & 80.0 MB/s & 230 MB/s\\
    \hline
\end{tabularx}
\end{table}

\begin{figure}
    \centering
\input{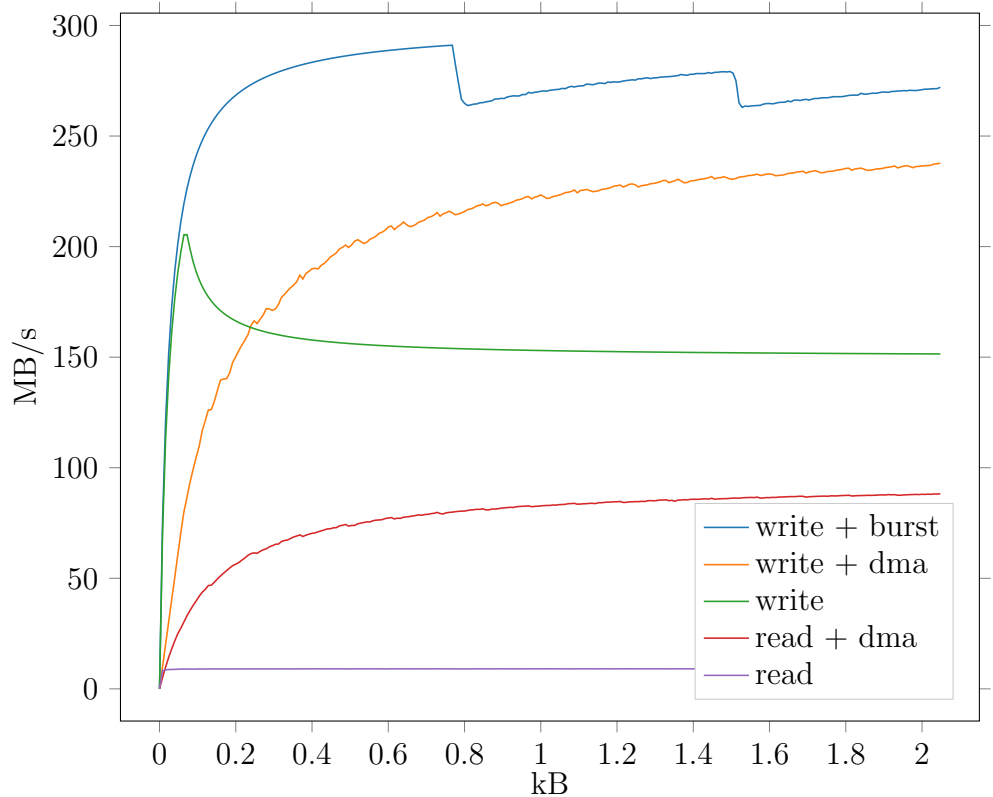}
    \caption{Different read and write speeds from a single core to external-memory when the network is free (no other cores are active). The horizontal axis shows the size of the data that was being written or read and the vertical axis shows the speed in MB/s. Because there is a small overhead associated with reading or writing to external memory the speeds are slow for very small sizes. Burst refers to hardware support for faster memory writes that is activated when consecutive 8-byte writes are performed. The non-burst writes are to non-consective locations. A possible explanation for the jumps in the blue line (write + burst) is that the burst mode gets interrupted after a specific number of bytes have been written. The non-monotonic behaviour of the green line (write) is due to a buffering effect of the Epiphany network mesh. }
    \label{fig:writespeed}
\end{figure}

We summarize the results of a number of measurements of the memory speed of the Parallella that we performed in Table \ref{tab:benchmarkparallella}. In Figure \ref{fig:writespeed} we show the results of one particular such measurement, regarding the reading and writing to external memory in a non-contested network state. From these results we can estimate $e$. Note that there is a significant difference between the read and write speeds when multiple cores are communicating with the external memory at the same time. We will choose to use the most pessimistic number, the read speed using the DMA engine from the external memory with a contested network state, since we expect that all cores will simultaneously be reading from the external memory during a hyperstep. We have found experimentally that a core of the Epiphany-III chip is on average performs the equivalent of one FLOP per 5 clock cycles in representative programs implementing BSPS algorithms that are compiled using GCC 4.8.2. We do note however that with hand-optimized assembly code as many as the equivalent of two FLOPs per clock cycle can be performed when performing multiplications and additions in succession, but we will not make use of this to preserve generality, so that the values we present are valid for real world BSPS algorithms implemented on top of Epiphany BSP. We then find that the external inverse bandwidth for this platform is:
$$e \approx (11~\mathrm{MB/s})^{-1} \approx 43.4 ~\mathrm{FLOP/float},$$
where we used that an Epiphany core runs at a default frequency of $600~\mathrm{MHz}$. Also we use single-precision floats which have a size of $4$ bytes on this platform. Note that from a practical perspective, this value for $e$ is sometimes prohibitively high, which means that we need to perform a large number of FLOPs with every floating point number we obtain or the time of a hyperstep will have the bandwidth as a bottleneck. This is an obvious limitation of the Parallella board, and is specific to this computer. We note that this high value for $e$ is not a general property of the Epiphany chip (nor any other BSP accelerator).
For $g$ and $l$ we fit a linear function against the raw measurements that were obtained for core-to-core writes for a varying number of bytes. We note that the Epiphany hardware is such that this specific type of communication does not suffer from the large discrepancies between simultaneous and non-simultaneous communication of multiple cores. After compensating for overhead because of the hardware clock that was used to perform the measurements\footnote{Starting and stopping the hardware clock takes a specific, fixed number of clock cycles. This offset has been subtracted from our measurements.}, we obtain for the barrier time (the latency):
$$l \approx 136~\mathrm{FLOP},$$
and for the inverse bandwidth of inter-core communication:
$$g \approx 5.59~\mathrm{FLOP/float}.$$
\new{We note that this is an upper bound, because one can obtain a value of $g$ lower than $1~\mathrm{FLOP/float}$ when using only optimized writes instead of reads. Furthermore, the startup cost of inter-core communication is less than one FLOP so the value of $l$ is almost entirely due to the synchronization mechanism and not due to starting up communication.}

\section{Experimental results}
\label{sec:results}

\subsection{Cannon's algorithm}
\label{sec:cannonresults}
\begin{figure}
    \centering
    \includegraphics[width=\linewidth]{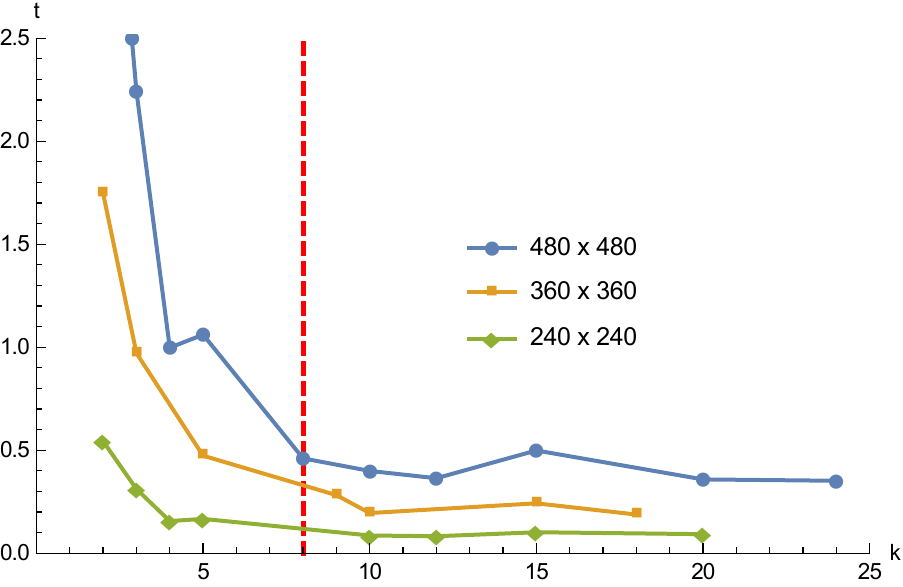}
    \caption{Run time of Cannon's algorithm on the Epiphany-III processor. The different lines correspond to different sizes of the matrix, as indicated by the legends. We show the value of $k = \frac{n}{NM}$ defined before, and the time in seconds that it took to run the algorithm. The value for $k_{\text{equal}}$ is shown using a red, dashed vertical line.}
    \label{fig:cannonbench}
\end{figure}

We have implemented Cannon's algorithm for dense matrix multiplication as discussed in Section \ref{sec:examples} and measured the running time for different parameters on the E16G301 chip found on the Parallella-16 micro-server with the Zynq 7010 SOC. The benchmark, which is part of the Epiphany BSP library, was compiled with GCC 4.8.2 using the Epiphany SDK version 2016.3. The results are shown in Figure \ref{fig:cannonbench}. The BSPS cost function of the algorithm, Equation \ref{eq:cannontime}, shows that the number of FLOPs required for computing the multiplication does not depend on $M$ (first term) but the communication volume (second and third term) does scale with $M$. Indeed, we expect a higher value of $M$, which results in a \emph{smaller} block size, to give a higher run time and this is in agreement with the results in Figure \ref{fig:cannonbench}. The block size should always be chosen as large as the limited amount of local memory allows.

Equating the left and right hand side of Equation \ref{eq:cannontime}, and solving for $k$ using the values we have found for the Epiphany processor yields $k_{\text{equal}} \approx 8$. Here, $k_{\text{equal}}$ corresponds to the boundary values for computation heavy and communication heavy hypersteps. As shown by Figure \ref{fig:cannonbench} this corresponds to the transition of communication heavy hypersteps to computation heavy hypersteps, and this is verified by our experiments. This shows that the BSPS cost function is a good way to identify possible bottlenecks for a BSPS algorithm, as well as being able to predict its running time.

\section{Future work}

There is still a wide range of algorithms in e.g. numerical linear algebra, scientific computing or computational geometry that we have not considered in this context but for which there exist efficient algorithms within the BSP model. We have some preliminary work on sparse matrix vector multiplication and external sorting within the BSPS model.

Furthermore, there are many real-world applications to be explored. As an example, we imagine applying the BSPS cost function to real-time video processing, where a frame is analyzed in each hyperstep. Here we could require the hypersteps to be bandwidth heavy to ensure that we are able to process the entire video feed in real-time.

While we focus on many-core coprocessors in this article, the same principles hold for any type of hardware that has to process data that is too large to fit in working memory. Therefore, these streaming algorithms may also be applied in \emph{Big Data} contexts.

Finally, \new{it would be interesting to} consider models in which there are different types of processing units, and to develop models that uses the BSP and BSPS costs to distribute the work of a single algorithm in this heterogeneous environment.

\section{Acknowledgements}

We would like to thank Rob Bisseling for many useful discussions. Also, we are grateful to Adapteva for providing part of the hardware that was used for the development of the software that was introduced in this paper, allowing us to perform the measurements that were presented.



\section*{References}
\bibliographystyle{elsarticle-num}
\bibliography{mendeley}

\begin{thebibliography}{10}
\expandafter\ifx\csname url\endcsname\relax
  \def\url#1{\texttt{#1}}\fi
\expandafter\ifx\csname urlprefix\endcsname\relax\def\urlprefix{URL }\fi
\expandafter\ifx\csname href\endcsname\relax
  \def\href#1#2{#2} \def\path#1{#1}\fi

\bibitem{Valiant1990}
L.~G. . V.~. Valiant, {A Bridging Model for Parallel Computation},
  Communications of the ACM 33~(8) (1990) 103--111.
\newblock \href {http://dx.doi.org/10.1145/79173.79181}
  {\path{doi:10.1145/79173.79181}}.

\bibitem{Alon1999}
N.~Alon, Y.~Matias, M.~Szegedy,
  \href{http://www.sciencedirect.com/science/article/pii/S0022000097915452}{{The
  Space Complexity of Approximating the Frequency Moments}}, Journal of
  Computer and System Sciences 58~(1) (1999) 137--147.
\newblock \href {http://arxiv.org/abs/arXiv:1505.00113v1}
  {\path{arXiv:arXiv:1505.00113v1}}, \href
  {http://dx.doi.org/10.1006/jcss.1997.1545}
  {\path{doi:10.1006/jcss.1997.1545}}.
\newline\urlprefix\url{http://www.sciencedirect.com/science/article/pii/S0022000097915452}

\bibitem{Babcock2002}
B.~Babcock, S.~Babu, M.~Datar, R.~Motwani, J.~Widom,
  \href{http://portal.acm.org/citation.cfm?doid=543613.543615}{{Models and
  issues in data stream systems}}, in: Proceedings of the Twenty-first ACM
  SIGMOD-SIGACT-SIGART Symposium on Principles of Database Systems, 2002, pp.
  1--16.
\newblock \href {http://dx.doi.org/10.1145/543614.543615}
  {\path{doi:10.1145/543614.543615}}.
\newline\urlprefix\url{http://portal.acm.org/citation.cfm?doid=543613.543615}

\bibitem{TOP500}
TOP500, \href{https://www.top500.org/lists/2016/06/}{{The List. June 2016.}}
\newline\urlprefix\url{https://www.top500.org/lists/2016/06/}

\bibitem{Intel}
Intel{\textregistered},
  \href{https://software.intel.com/en-us/xeon-phi/mic}{{Xeon Phi™
  Coprocessor}}.
\newline\urlprefix\url{https://software.intel.com/en-us/xeon-phi/mic}

\bibitem{Olofsson2015}
A.~Olofsson, T.~Nordstr{\"{o}}m, Z.~Ul-Abdin, {Kickstarting high-performance
  energy-efficient manycore architectures with Epiphany}, in: Conference Record
  - Asilomar Conference on Signals, Systems and Computers, Vol. 2015-April,
  2015, pp. 1719--1726.
\newblock \href {http://arxiv.org/abs/arXiv:1412.5538}
  {\path{arXiv:arXiv:1412.5538}}, \href
  {http://dx.doi.org/10.1109/ACSSC.2014.7094761}
  {\path{doi:10.1109/ACSSC.2014.7094761}}.

\bibitem{Movidius}
Movidius,
  \href{http://www.movidius.com/solutions/vision-processing-unit}{{Myriad 2
  Vision Processor Unit}}.
\newline\urlprefix\url{http://www.movidius.com/solutions/vision-processing-unit}

\bibitem{Kalray}
Kalray,
  \href{http://www.kalrayinc.com/kalray/products/{\#}processors}{{MPPA{\textregistered}:
  The Supercomputing on a chip™ solution}}.
\newline\urlprefix\url{http://www.kalrayinc.com/kalray/products/{\#}processors}

\bibitem{McColl1999}
W.~F. McColl, a.~Tiskin, {Memory-Efficient Matrix Multiplication in the BSP
  Model}, Algorithmica 24~(3-4) (1999) 287--297.
\newblock \href {http://dx.doi.org/10.1007/PL00008264}
  {\path{doi:10.1007/PL00008264}}.

\bibitem{Tiskin1998}
A.~Tiskin,
  \href{http://linkinghub.elsevier.com/retrieve/pii/S0304-3975(97)00197-7}{{The
  bulk-synchronous parallel random access machine}}, Theoretical Computer
  Science 196~(1-2) (1998) 109--130.
\newblock \href {http://dx.doi.org/10.1016/S0304-3975(97)00197-7}
  {\path{doi:10.1016/S0304-3975(97)00197-7}}.
\newline\urlprefix\url{http://linkinghub.elsevier.com/retrieve/pii/S0304-3975(97)00197-7}

\bibitem{Dehne2003}
F.~Dehne, W.~Dittrich, D.~Hutchinson, {Efficient External Memory Algorithms by
  Simulating Coarse-Grained Parallel Algorithms} (2003) 97--122.

\bibitem{Dehne2002}
F.~Dehne, W.~Dittrich, D.~Hutchinson, A.~Maheshwari, {Bulk synchronous parallel
  algorithms for the external memory model}, Theory of Computing Systems 35~(6)
  (2002) 567--597.
\newblock \href {http://dx.doi.org/10.1007/s00224-002-1066-2}
  {\path{doi:10.1007/s00224-002-1066-2}}.

\bibitem{Gerbessiotis2015}
A.~V. Gerbessiotis,
  \href{http://dx.doi.org/10.1016/j.parco.2014.12.002}{{Extending the BSP model
  for multi-core and out-of-core computing: MBSP}}, Parallel Computing 41
  (2015) 90--102.
\newblock \href {http://dx.doi.org/10.1016/j.parco.2014.12.002}
  {\path{doi:10.1016/j.parco.2014.12.002}}.
\newline\urlprefix\url{http://dx.doi.org/10.1016/j.parco.2014.12.002}

\bibitem{Valiant2011}
L.~G. Valiant, {A bridging model for multi-core computing}, Journal of Computer
  and System Sciences 77~(1) (2011) 154--166.
\newblock \href {http://dx.doi.org/10.1016/j.jcss.2010.06.012}
  {\path{doi:10.1016/j.jcss.2010.06.012}}.

\bibitem{Bilardi2001}
G.~Bilardi, C.~Fantozzi, A.~A. Pietracaprina, G.~Pucci,
  \href{http://www.springerlink.com/content/re5mk4uu2596j56c/}{{On the
  Effectiveness of D-BSP as a Bridging Model of Parallel Computation}},
  International Conference on Computational Science (2001) 579--588\href
  {http://dx.doi.org/10.1007/3-540-45718-6} {\path{doi:10.1007/3-540-45718-6}}.
\newline\urlprefix\url{http://www.springerlink.com/content/re5mk4uu2596j56c/}

\bibitem{Cannon1969}
L.~E. Cannon, {A Cellular Computer to Implement the Kalman Filter Algorithm
  (No. 603-Tl-0769).}

\bibitem{Ross2015}
J.~A. Ross, D.~A. Richie, S.~J. Park, D.~R. Shires, {Parallel programming model
  for the Epiphany many-core coprocessor using threaded MPI} (2015).
\newblock \href {http://dx.doi.org/10.1016/j.micpro.2016.02.006}
  {\path{doi:10.1016/j.micpro.2016.02.006}}.

\bibitem{Hill1998}
J.~M. Hill, B.~McColl, D.~C. Stefanescu, M.~W. Goudreau, K.~Lang, S.~B. Rao,
  T.~Suel, T.~Tsantilas, R.~H. Bisseling,
  \href{http://linkinghub.elsevier.com/retrieve/pii/S0167819198000933}{{BSPlib:
  The BSP programming library}}, Parallel Computing 24~(14) (1998) 1947--1980.
\newblock \href {http://dx.doi.org/10.1016/S0167-8191(98)00093-3}
  {\path{doi:10.1016/S0167-8191(98)00093-3}}.
\newline\urlprefix\url{http://linkinghub.elsevier.com/retrieve/pii/S0167819198000933}

\bibitem{Yzelman2014}
A.~N. Yzelman, R.~H. Bisseling, D.~Roose, K.~Meerbergen, {MulticoreBSP for C: A
  high-performance library for shared-memory parallel programming},
  International Journal of Parallel Programming\href
  {http://dx.doi.org/10.1007/s10766-013-0262-9}
  {\path{doi:10.1007/s10766-013-0262-9}}.

\bibitem{Buurlage2015}
J.-W. Buurlage, T.~Bannink, A.~Wits, \href{http://www.codu.in/ebsp/}{{Epiphany
  BSP 1.0}} (2015).
\newline\urlprefix\url{http://www.codu.in/ebsp/}

\bibitem{Adapteva2013}
Adapteva,
  \href{http://www.adapteva.com/docs/epiphany{\_}arch{\_}ref.pdf}{{Epiphany
  Architecture Reference}}.
\newline\urlprefix\url{http://www.adapteva.com/docs/epiphany{\_}arch{\_}ref.pdf}

\bibitem{Olofsson}
A.~Olofsson, {Epiphany-V: A 1024 processor 64-bit RISC System-On-Chip}  1--15.

\bibitem{Agathos2015}
S.~N. Agathos, A.~Papadogiannakis, V.~V. Dimakopoulos,
  \href{http://dx.doi.org/10.1007/978-3-662-48096-0{\_}51}{{Targeting the
  Parallella}}, in: L.~J. Tr{\"{a}}ff, S.~Hunold, F.~Versaci (Eds.), Euro-Par
  2015: Parallel Processing: 21st International Conference on Parallel and
  Distributed Computing, Vienna, Austria, August 24-28, 2015, Proceedings,
  Springer Berlin Heidelberg, Berlin, Heidelberg, 2015, pp. 662--674.
\newblock \href {http://dx.doi.org/10.1007/978-3-662-48096-0}
  {\path{doi:10.1007/978-3-662-48096-0}}.
\newline\urlprefix\url{http://dx.doi.org/10.1007/978-3-662-48096-0{\_}51}

\bibitem{Fleming2015}
M.~Fleming,
  \href{https://www.parallella.org/2015/03/20/erlang-otp-and-the-parallella-board/}{{Erlang-OTP
  and the Parallella Board}} (2015).
\newline\urlprefix\url{https://www.parallella.org/2015/03/20/erlang-otp-and-the-parallella-board/}

\bibitem{BrownDeerTechnology2014}
{Brown Deer Technology LLC},
  \href{http://www.browndeertechnology.com/coprthr.htm}{{COPRTHR{\textregistered}
  (CO-PRocessing THReads{\textregistered})}} (2014).
\newline\urlprefix\url{http://www.browndeertechnology.com/coprthr.htm}

\end{thebibliography}





\end{document}